\documentclass[aps,pre,reprint,twocolumn,doi=false,isbn=false,url=false,title=true,longbibliography,noeprint,footinbib]{revtex4-1}

\usepackage{array}
\usepackage{amsfonts}
\usepackage{float}
\usepackage{xcolor}
\usepackage{amsmath}
\usepackage{amssymb}
\usepackage{graphicx,color,soul}
\usepackage{csquotes}

\def\ba{\begin{eqnarray}}
\def\ea{\end{eqnarray}}
\def\be{\begin{equation}}
\def\ee{\end{equation}}
\def\bm{\begin{math}}
\def\me{\end{math}}

\newcommand{\mean}[1]{\langle #1\rangle}

\makeatletter
\newcommand{\fmarki}{*}
\newcommand{\fmarkii}{\ensuremath{\dagger}}
\newcommand{\fmarkiii}{\ensuremath{\ddagger}}
\newcommand{\fmarkiv}{\ensuremath{\mathsection}}
\newcommand{\fmarkv}{\ensuremath{\mathparagraph}}
\newcommand{\fmarkvi}{\ensuremath{\|}}
\newcommand{\fmarkvii}{**}
\newcommand{\fmarkviii}{\ensuremath{\dagger\dagger}}
\newcommand{\fmarkix}{\ensuremath{\ddagger\ddagger}}
                
\def\@fnsymbol#1{{\ifcase#1\or \fmarki\or \fmarkii\or \fmarkiii\or \fmarkiv\or \fmarkv\or \fmarkvi\or \fmarkvii\or \fmarkviii\or \fmarkix \else\@ctrerr\fi}}
\makeatother

\renewcommand{\fmarki}{$\dagger$}
 \renewcommand{\fmarkii}{*}
 \renewcommand{\fmarkiii}{$\ddagger$}
 \renewcommand{\fmarkiv}{a$_4$}
 \renewcommand{\fmarkv}{x$_5$}

\begin{document}

\title{{Scale-free cluster-cluster aggregation during polymer collapse}}

\author{Suman Majumder}\email[]{smajumder.@amity.edu, suman.jdv@gmail.com}

\affiliation{Amity Institute of Applied Sciences, Amity University Uttar Pradesh, Noida 201313, India}

\author{Saikat Chakraborty}\email{saikat.chakraborty@univ-grenoble-alpes.fr}
\affiliation{Laboratoire Interdisciplinaire de Physique, Universit\'{e} Grenoble Alpes, CNRS, 38402 Saint-Martin-d'H\'{e}res, France}


\begin{abstract}
{\bf Abstract:} An extended polymer collapses to form a globule when subjected to a quench below the collapse transition temperature. The process begins with the formation of clusters of monomers or ``pearls''. The nascent clusters merge, resulting in growth of the average cluster size $C_s$, eventually leading to a single globule. The aggregation of the clusters are known to be analogous to droplet coalescence. This suggests a striking resemblance between such an aggregation  and cluster-cluster aggregation found in many {particle systems}, like in colloidal self-assembly, typically characterized by a universal dynamic scaling behavior. Motivated by that, here, we verify the presence of such  dynamic scaling during the collapse of a polymer with varying bending stiffness $\kappa$, using molecular dynamic simulations.  We probe the dynamics via time evolution of the size distribution of clusters $N_s(t)$ and growth of $C_s(t)$.  Irrespective  of $\kappa$, we observe the power-law scalings $C_s(t)\sim t^z$ and $N_s(t)\sim t^{-w} s^{-\tau}$, of which only the cluster growth is universal with {$z\approx 1.67$.} Importantly, our results indeed show that $N_s(t)$ exhibits a dynamic scaling of the form $N_s(t)\sim s^{-2}f(s/t^z)$, indicative of a scale-free cluster growth. Interestingly, for flexible and weakly stiff polymers the dynamic exponents obey the relation $w=2z$, as also found in diffusion-controlled cluster-cluster aggregation of particles.  For $\kappa \ge 5$, the exponents show deviation from this relation, which grows  continuously with $\kappa$.  We identify the differences in local structures of the clusters formed, leading to  variations in cluster-size dependence of the effective diffusion constant to be the origin of the above deviation. We also discuss potential experimental strategies to directly visualize the observed dynamic scaling in a collapsing polymer.
\end{abstract}
\maketitle

\section{Introduction}
An extended polymer collapses to a globule when the monomer-monomer interaction overpowers the monomer-solvent interaction \cite{flory1953principles,stockmayer1960problems,nishio1979first}. Typically, this is achieved via a quench below the associated collapse transition temperature or $\Theta$-temperature. Following the quench, the polymer attains its equilibrium globular conformation through a nonequilibrium relaxation process \cite{degennes1985,byrne1995,timoshenko1995,kuznetsov1995,kuznetsov1996a,kuznetsov1996b,klushin1998,kikuchi2005,ye2007,pham2008,guo2011,Majumder2015,majumder2017,christiansen2017,majumder2019,majumder2020}. Study of the nonequilibrium dynamics during the collapse have implications in understanding folding pathways of globular proteins \cite{hagen2000,Sadqi2003,haran2012,udgaonkar2013,reddy2017}. In this context, recently it has been shown that intrinsically disordered proteins can have multiple stable structures akin to coil or globules \cite{chakraborty2025,zeng2022}. Conformational switch between two states may be mapped to a coil-globule transition in polymers. Thus, a deeper understanding of the nonequilibrium process is still of much fundamental interest. 
\par
The study of the collapse of a polymer can be traced back to the work of de Gennes \cite{degennes1985}. It proposes the formation of dense ``sausage'' shaped conformations at early times through a uniform aggregation across the chain. Drive to minimize the interfacial free energy results in thickening of the sausage, eventually forming a spherical globule. Later studies suggest an alternate sequence of events, where dense regions along the polymer chain nucleate to form small clusters of monomers or ``pearls'', giving rise to the so called ``pearl-necklace'' intermediate. Subsequently, these pearls grow further by acquiring monomers from adjacent segments of the polymer and at later they self aggregate upon collision forming a single cluster or globule. This progression, proposed by Halperin and Goldbart \cite{halperin2000} is slightly different from that of Klushin  \cite{klushin1998}. A chain-tension mediated merging of the clusters to final globule state is put forward in the latter's work. Since monitoring a single polymer was difficult, experimental verification of this multi-stage collapse was rare in the past \cite{xu2006,ye2007}, and hence most studies relied on computer simulations. Almost all of these studies confirmed the ``pearl-necklace'' as the phenomenological picture \cite{majumder2020}. As a matter of fact, there are even evidences of observation of ``pearl-necklace'' in experiments as well \cite{ye2007}. However, recently it has been shown that depending on temperature and solvent viscosity the collapse of a flexible homopolymer may proceed through a combination of sausage and pearl-necklace intermediates \cite{majumder2024temperature}.
\par
Focus of the existing studies, mostly using flexible homopolymers, was on extracting the scaling of the collapse time with the length of the polymer and growth kinetics of the ``pearls'' or clusters. In this regard, only recently an analogy has been drawn between this cluster growth with the usual coarsening phenomena in spin or particle system \cite{bray1994theory}. Consequently, various tools from the well-established field of coarsening phenomena have been successfully employed in understanding the collapse kinetics of a flexible homopolymer \cite{Majumder2015,majumder2017,christiansen2017,majumder2019,majumder2020}. Even though this allowed  numerical estimations of the scaling of the collapse time and cluster growth, a unified understanding for different classes of polymers (i.e., for semiflexible polymers) based on pre-existing theoretical foundations is still lacking. 
\par
Cluster-cluster aggregation is the underlying mechanism of self-assembly across a range of length scale from cosmology \cite{hawking1982bubble} to intra-cellular organization \cite{lee2023size}. In this regard, the work of Smoluchowski  is the first to provide a mean-field theory of clustering \cite{smoluchowski1916}. Later, based on the seminal diffusion-limited aggregation model \cite{witten1981diffusion}, more general models for kinetics of cluster-cluster aggregation have been proposed \cite{meakin1983formation,kolb1983scaling}, which can effectively overcome the limitations of the mean-field description. In this context, Vicsek and Family \cite{vicsek1984} suggested that the cluster size distributions as a function of cluster size and time follow a universal dynamic scaling form  \cite{family1985cluster,miyazima1987,lin1989universality,lin1990universal}. This has been later verified in numerous systems exhibiting cluster-cluster aggregation. In particular, experiments on colloidal aggregates have established that the underlying scaling functions are universal, being independent of the chemical composition of the colloids \cite{weitz1984dynamics,weitz1986dynamic,lin1989universality,valadez2018reversible}. Apart from colloids, experiments in externally driven ferrofluids, cellular organelles, or surface aggregations can also be successfully described by the theory \cite{vcernak2004,shahrivar2017,lee2023size}. 
\par 
On the other hand, as already mentioned there have been multiple studies that indicate the presence of cluster-cluster aggregation during the collapse of a polymer. In spite of that, to the best of our knowledge, no effort has been dedicated to apply the  Vicsek-Family scaling principles in understanding the same. The necessity of such an approach, advocating universality in cluster sizes is increasingly becoming important as conformational changes in complex bio-macromolecule like proteins require better understanding of the underlying principles of polymer physics. To this end, here, {by means of molecular dynamics} (MD) simulations we investigate the kinetics of collapse of homopolymers, ranging from flexible to moderately stiff ones (semiflexible), with an emphasis on exploring the presence of any dynamic scaling, as found in colloidal systems.  
\par
The rest of the paper is organized in the following manner. Next, in Sec.\ \ref{model} we introduce the model polymer and details of the performed MD simulations. Following that in Sec.\ \ref{theory} we provide a brief description of the existing scaling theory for particle systems, and thereby introduce the observables which we calculate. In Sec.\ \ref{results} we present the simulation results. Finally, we present summary and discussions in Sec.\ \ref{conclusion}.

\section{Model and Methods}\label{model}
We consider a coarse-grained bead-spring model polymer chain consisting of $N$ monomers or beads of mass $m$ and diameter $\sigma$, connected in a linear fashion.  The bond between successive monomers, which are at a distance $r$ apart is realized via the standard finitely extensible non-linear elastic (FENE) potential \cite{kremer1990dynamics}

\begin{equation}
V_{\text{FENE}}(r) = - \frac{1}{2} K R_0^2 \ln \left[1 - \left(\frac{r}{R_0}\right)^2\right]; ~r<R_0,
\end{equation}
where we chose the spring constant $K= 30\epsilon/\sigma^2$ and $R_0 = 1.5 \sigma$. The monomers also interact  with each other via the standard Lennard-Jones (LJ) potential
\begin{equation}
V_{\text{LJ}}(r) = 4 \epsilon \left[ \left( \frac{\sigma}{r} \right)^{12} - \left( \frac{\sigma}{r} \right)^{6} \right], 
\end{equation}
with the interaction strength $\epsilon$. The repulsive part in $ V_{\text{LJ}}$ takes care of the volume exclusion of the monomers and the attractive part ensures a collapse transition for the polymer as a function of temperature. For convenience the LJ potential is truncated, shifted and force corrected at $r_c = 2.5 \sigma$, leading to 
\begin{equation}\label{nb_poten}
  V_{\rm{LJ}}^{\rm mod}(r)=
\begin{cases}
  V_{\rm{LJ}}(r)-V_{\rm{LJ}}(r_c) -(r-r_c)\frac{dV_{\rm{LJ}}}{dr}\Big|_{r=r_c}  r<r_c \,,\\
0 ~~~~~~~~~ \text{otherwise}\,.
   \end{cases}
\end{equation}
The modified potential behaves {qualitatively} the same as $V_{\text{LJ}}$, however, is continuous and differentiable at $r=r_c$. The energy penalty for bending of the polymer backbone is given by 
\begin{equation}
V_{\text{bend}} = \kappa \sum_{i=1}^{N-2} \left( 1 - \cos \theta_i \right),
\end{equation}
where $\theta_i$ is the angle between consecutive bonds and $\kappa$ is the bending stiffness. The case of $\kappa=0$ stands for a flexible polymer with no bending energy penalty and $\kappa > 0$ corresponds to a semiflexible polymer. The described model exhibits a collapse transition as a function of temperature where at high temperature the polymer attains an extended coil conformation and at low temperature it remains as a compact conformation. 
\par

We perform MD simulations of the above model. The equation of motions of the polymer beads are solved using the standard velocity-Verlet integration scheme \cite{frenkel_book}. The temperature is kept constant using the Nos\'{e}-Hoover chain thermostat \cite{nose1984unified,hoover1985canonical,martyna1992nose}. The discrete time step of integration is chosen to be $\Delta t = 5 \times 10^{-4} \tau_0$, where $\tau_0= \sqrt{m\sigma^2/\epsilon}$ is the standard LJ unit of time. The unit of temperature $T$ is $\epsilon/k_B$. For convenience, in our simulations we set $m$, $\sigma$ and $\epsilon$ to unity. For faster computation we rely on the simulation package LAMMPS \cite{plimpton1995}. 
\par
Since we study the nonequilibrium dynamics of the collapse transition of a polymer, an extended-coil conformation of a polymer is desired as the initial configuration. For that we use a self-avoiding walk of length $N-1$ as a polymer with $N$ monomers. {Using this self-avoiding walk we first perform MD simulation at a high $T$ for a period $\tau_{\rm eq}\sim N^2$ such that the polymer chain attains an extended-coil conformation. Following that we start the nonequilibrium evolution by setting $T=1$.  As we are interested in the kinetics of collapse transition, here, we present result for polymer with moderate bending stiffness, i.e., $\kappa \in [0,10]$. For even larger $\kappa$, the collapse transition accompanies with other complex conformational transitions \cite{majumder2021knots}. Majority of the results are based on polymers of  chain lengths, $N\le 2048$. However, for further confirmation, for three specific values of $\kappa$, we present results from relatively much longer chain length of $N=8192$. The equilibration time $\tau_{\rm eq}$ at high $T$ for $N=512$ is $100 \tau_0$. For other chain lengths, $\tau_{\rm eq}$ is chosen using the relation $\tau_{\rm eq}\sim N^2$. Following the quench, we simulated the chain until it collapses to a single globule. The typical timescale for the collapse is $\tau_c \sim N$ in MD simulations \cite{majumder2020}. In this study, as we will see later that the collapse time is also depended on the polymer stiffness. Typical maximum run time chosen for a chain of length $N=2048$ varies from $200 \tau_0$ to $800 \tau_0$ as $\kappa$ varies from $0$ to $10$.}  Except the evolution snapshots and contact maps all data presented are averaged over $500$ independent realizations obtained by using different seed for the random number while generating the self-avoiding walk.

\section{Dynamic scaling theory}\label{theory}
Before we move forward to the simulation results, here, we briefly discuss the dynamic scaling theory pertinent to understand cluster-cluster aggregation during colloidal self assembly. During the course of this discussion we also introduce various relevant quantities that will be used later. Here, by a cluster of size $s$ we refer to the number of particles (or monomers) forming the cluster. In our cluster identification method,  an aggregate of monomers with $s$ beyond a threshold $s_c=10$ is considered to be a cluster~\cite{majumder2020}.  Thus, if $N_s(t)$ is the number of clusters of size $s$ at time $t$,  the total number of clusters  is 
\begin{equation}
 n_c(t)=\sum_{s=10}N_s(t),
\end{equation}
 and the average size of the clusters 
\begin{equation}\label{eq:clssizedef}
C_s(t) = \frac{\sum_{s} {sN_s(t)}}{n_c(t)}. 
\end{equation}
Typically, cluster-cluster aggregation occurs via a power-law growth as  
\begin{equation}\label{eq:clssize}
C_s(t) \sim t^z,
\end{equation}
where $z$ is the corresponding growth exponent. The dynamic scaling theory relates the cluster size distribution to their time evolution through the expression \cite{vicsek1984} 
\begin{equation}\label{eq:fvnomass}
   N_s(t) \sim s^{-2} f\left(\frac{s}{t^z} \right). 
\end{equation}
The scaling function $f(x)$ exhibits a power-law behavior for $x \ll 1$: $f(x) \sim x^\delta$, where $\delta$ is the crossover exponent. For very large $x$, one gets $f(x) \ll 1$. 
Thus, at a fixed $t$ when $s \ll C_s(t)$, one expects a power-law dependence of $N_s(t)$ on $s$ and $t$ as 
\begin{equation}\label{eq:fvstscale}
    N_s(t) \sim s^{-\tau} t^{-w},
\end{equation} 
with  $w=z\delta$ and $\tau=2-\delta$. This leads to the scaling relation
\begin{equation}\label{eq:sclnomass}
    w = (2-\tau)z.
\end{equation}
\par
Alternatively, $N_s(t)$ may have a different relation with $s$ and $t$, given as \cite{meakin1985}
\begin{equation}\label{eq:fvmass}
    N_s(t) \sim s^{-2} \Tilde{f}\left(\frac{s}{t^z} \right),
\end{equation}
where $\Tilde{f} = x^2g(x)$ is a different scaling function. The cut-off function $g(x)$ describes a fast decay for both $x \gg 1$ and $\ll 1$. In this case, $N_s(t)$ will have bell-shaped profiles. Thus, rewriting the $s$ and $t$ dependence similar to Eq.~\eqref{eq:fvstscale}, one arrives at a different scaling relation given by 
\begin{equation}\label{eq:sclmass}
    w=2z.
\end{equation}
The dynamic scaling theory was generalized \cite{meakin1985} by showing its dependence on the diffusion of the clusters. The diffusion constant $D_s$ of cluster of size $s$ scales as 
\begin{equation}\label{D_eq}
 D_s \sim s^\gamma,
\end{equation}
where the exponent $\gamma$ decides the type scaling relation. The generalized theory predicts that for $\gamma > \gamma_c$ the relation \eqref{eq:sclnomass} is valid and for $\gamma < \gamma_c$ one expects 
Eq.~\eqref{eq:sclmass} to be applicable. Here, $\gamma_c$ is a critical value below which aggregation dynamics is dominated only by the large cluster–small cluster interactions, and  above which large-small and large-large clusters processes contribute equally. Thus, combining the two cases one can write down a general scaling relation 
\begin{equation}\label{delta}
 w=z\delta,
\end{equation}
where the crossover exponent $\delta$ is given by
\begin{equation}
 \delta=
\begin{cases}
  (2-\tau)~~~ \text{for}~ \gamma> \gamma_c \\
  ~~~~~2~~~~~~~~\text{for}~ \gamma < \gamma_c\,.
   \end{cases}
\end{equation}

\par
{Origin of the dynamic scaling of the form \eqref{eq:fvnomass} or \eqref{eq:fvmass} was purely empirical \cite{vicsek1984,meakin1985}. The scaling relation in Eq.\ \eqref{delta} is purely a consequence of the constant density of the system. Thus, even though it was initially realized in diffusion-limited aggregation of clusters such as during  colloidal self-assembly, over the years it has been observed that systems with different transport mechanisms too exhibit such dynamic scaling. This includes interface growth during roughening transition, active Eden process, growth due to stochastic ballistic deposition and even interface growth in turbulent liquid crystals \cite{family1985scaling,plischke1985dynamic,kim1989growth,maunuksela1997kinetic,myllys2001kinetic,takeuchi2010universal}.}


\begin{figure*}
    \centering
    \includegraphics[width=6.0in]{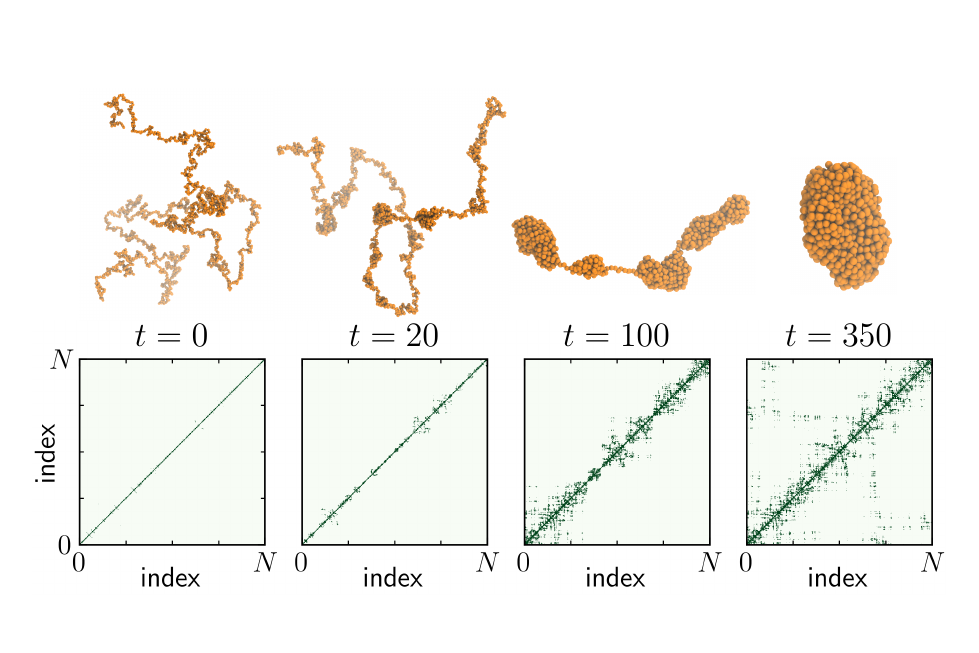}
    \caption{Collapse of a flexible polymer. Representative snapshots at different times and their corresponding contact maps illustrating the sequence of events during the collapse of a flexible polymer of length $N=2048$. The first frame shows the nucleation of small clusters or ``pearl'' along the polymer chain, indicated by development of monomer-monomer contacts (dark blue region). The second and third frames reflect growth of clusters with long (contour) distance contacts. The final frame represents the globular state with contacts spanning the whole chain.}
    \label{fig:mapflex}
\end{figure*}
\begin{figure}[b!]
    \centering
    \includegraphics[width=3.0in]{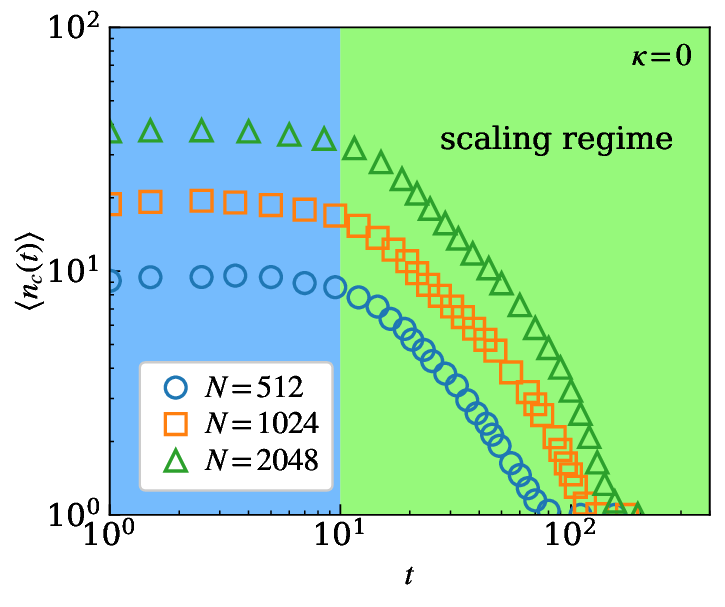}
    \caption{Cluster-cluster aggregation in flexible polymers. Time evolution of the average number of clusters $\mean{n_c}$ for three different chain lengths. Profiles of $\mean{n_c}$ show a plateau at early times (blue background) followed by a monotonic decay (green background), consistent with the pear-necklace phenomenological picture.}
    \label{fig:clusflex}
\end{figure}
\section{Results}\label{results}

\subsection{Flexible polymers} \label{flexible}
We begin by showing representative conformations of a flexible polymer {of length $N=2048$ illustrating the time evolution during its collapse, in Fig.\ \ref{fig:mapflex}. Starting from an extended-coil-like conformation, at early times, clusters of monomers or ``pearls'' develop randomly along the chain. Such developments result in appearance of locally dense regions. To identify these clustered  regions on the chain, we estimate the inter-monomer pairwise distances for non-bonded monomers. We consider a pair of monomers to be in \textit{contact}, if their distance is less than $2.5\sigma$. On the two-dimensional contact maps, the colored diagonal corresponds to self contacts. These are almost the only contacts present at early times as shown in the first frame. Transient proximity of different segments of the fluctuating polymer generates the sporadic off-diagonal contacts.  The developments of clusters with time appear as few off-diagonal contacts. The second and the third frame depict such intermediate stage of collapse representing the ``pearl-necklace''. Visual inspection of the snapshot as the collapse progresses suggests decreased number of clusters, but with increased sizes . Consistent with the prediction of the pearl-necklace representation of the collapse, inspection of the trajectories reveal that the nucleated clusters merge with each other forming larger clusters. The corresponding contact map demonstrates this growth by increase in size of regions of off-diagonal or long-range contacts at the expense of their number. Finally, a single globule is formed after aggregation of all the clusters (final frame in Fig.~\ref{fig:mapflex}), resulting in appearance of contacts throughout the chain.

\par
To quantify these stages of collapse, we present the time evolution of the average number of clusters $\mean{n_c}$ for three different chain lengths in Fig.~\ref{fig:clusflex}. Here and later $\mean{\dots}$ indicates averaging over different simulation runs performed using independent initial conformations. Following the temperature quench, clusters nucleate fast, resulting in an almost instantaneous rise of $\mean{n_c}$. Afterwards, for a short time period spanning up to about $t=10$, the profiles remain almost flat. The plots are shifted upwards for longer polymers. A simultaneous inspection of the snapshots and contact maps in Figure~\ref{fig:mapflex} suggests that the clusters continue to grow during this time window. Clearly, at this stage the growth occurs by acquiring monomers from the chains connecting the clusters, having no impact on the number of clusters. The time span up to this stage appears to be independent of $N$.  A monotonic decay to unity or a collapsed globule state follows the plateau. Evidently, aggregation of smaller clusters result in such decrease of $\mean{n_c}$. As expected, the formation of globule takes longer for larger $N$. Here, we focus on the spatio-temporal scales, where $\mean{n_c}$ undergoes the monotonic decay at intermediate to late stages of the collapse, and identify this cluster-cluster aggregation dominated stage as our regime of interest.

\begin{figure}
    \centering
    \includegraphics[width=3.0in]{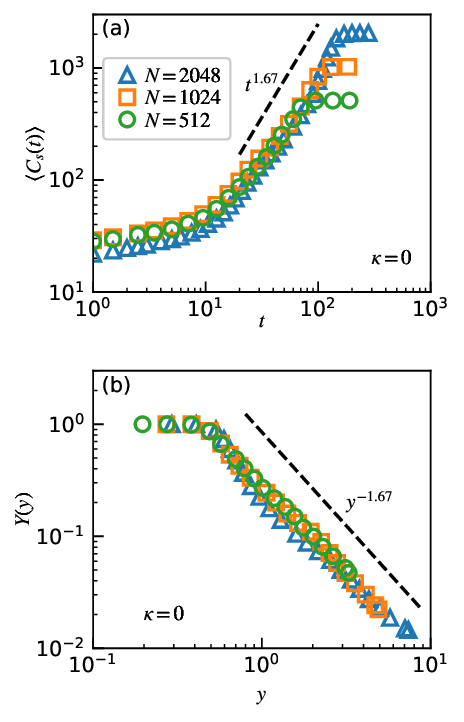}
    \caption{Cluster growth in flexible polymers. (a)~Growth of the average cluster size $\mean{C_s(t)}$ during the collapse of polymers of three different lengths $N$. The dashed line shows the consistency of the data with a power-law growth with an exponent $z=1.67$. (b)~Plots of the scaling functions $Y(y)$, demonstrating the collapse of data for different $N$ through a finite-size scaling analysis providing an unambiguous estimate of $z$. In both (a) and (b) the data are shown on a double-log scale.}
    \label{fig:clusgrowthflex}
\end{figure}
\par
After identifying the time regime of interest, we explore whether the dynamic scaling in Eq.~\eqref{eq:fvnomass} or \eqref{eq:fvmass} can be extended to aggregations of clusters of monomers under the topological constraint of a polymer chain. As the first step, we check the variation of the average cluster size with time following Eq.~\eqref{eq:clssizedef} [see Fig.~\ref{fig:clusgrowthflex}(a)]. The rise of $\mean{C_s(t)}$ for different $N$ follow similar profiles, barring the saturation at late times. Such deviations are due to effects of finite lengths of the polymer chains. The preceding similarity suggests presence of identical kinetics of growth. The sluggish growth of $\mean{n_c(t)}$ for small $t$ ($\lessapprox 10$) is a signature of monomer acquisition by the existing clusters or ``pearls''. This stage is followed by a faster growth which exhibits a linear behavior on the double-log scale, consistent with a power-law behavior in Eq.~\eqref{eq:clssize} with $z=1.67$. A comparison of time scales between Figs.~\ref{fig:clusflex} and \ref{fig:clusgrowthflex}(a) confirms that the time period when $\mean{C_s(t)}$ shows increase is indeed the cluster-cluster aggregation regime, i.e., our regime of interest.
\par
To obtain an unambiguous estimate of $z$, we account for the finite-size effects by constructing a finite-size scaling analysis \cite{majumder2017,majumder2018universal,schneider2020different}. For that we write down the following growth ansatz
\begin{equation}\label{growth-ansatz}
    \mean{C_s(t)} = C_0 + A_zt^z,
\end{equation}
where $C_0$ corresponds to the initial offset. To account for the finite-size effect we introduce a finite-size scaling function $Y(y)$ such that 
\begin{equation}
    Y(y) = \frac{\mean{C_s(t)}-C_0}{N-C_0}.
\end{equation}
{For the choice of the scaling variable
\begin{equation}
 y=\frac{(N-C_0)^{1/z}}{t},
\end{equation}
in this exercise by tuning the value $z$}, one thus expects that the data for different $N$ to collapse onto each other for the appropriate choice of $z$ along with a $Y(y) \sim y^{-z}$ behavior in the finite-size unaffected regime. The optimum result of this exercise is demonstrated in Fig.\ \ref{fig:clusgrowthflex}(b) that confirms the value of $z \approx 1.67$. 

\begin{figure}[t!]
    \centering
    \includegraphics[width=3.0in]{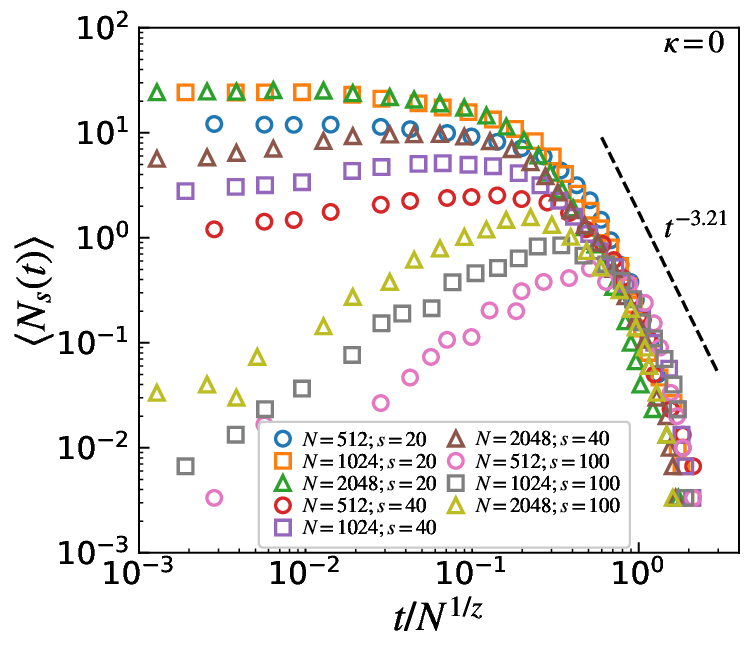}
    \caption{Kinetics of clusters of fixed sizes in flexible polymers.  Time dependence of the number $N_s$ of clusters of fixed size $s$ on a double-log scale, during the collapse. Plots for multiple $s$ across different $N$ are included. The abscissa is scaled with $N^{1/z}$ for a better visualization. $N_s(t)$ display both monotonic and bell-shaped profiles. The dashed line shows the consistency of the data with the power-law decay.}
    \label{fig:fixedclusgrowthflex}
\end{figure}
\begin{figure}[b!]
    \centering
    \includegraphics[width=3.0in]{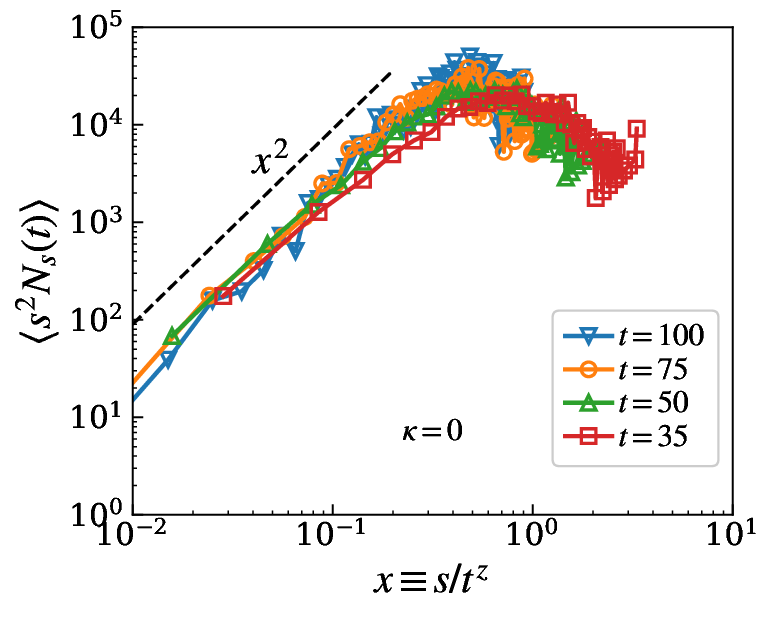}
    \caption{Verification of dynamic scaling in a flexible polymer. Double-log plots of the scaling function $f(x)\equiv s^2N_s(t)$ against the scaling variable $x\equiv s/t^z$ at different times for a polymer of length $N=2048$. The dashed black line shows the consistency of the data with the power law $\sim x^2$ in the regime $x\ll1$. Collapse of data at different times confirms the presence of colloid-like dynamic scaling during flexible polymer collapse.}
    \label{fig:dist_cluster_flex}
\end{figure}
\begin{figure*}[t!]
    \centering
    \includegraphics[width=6.0in]{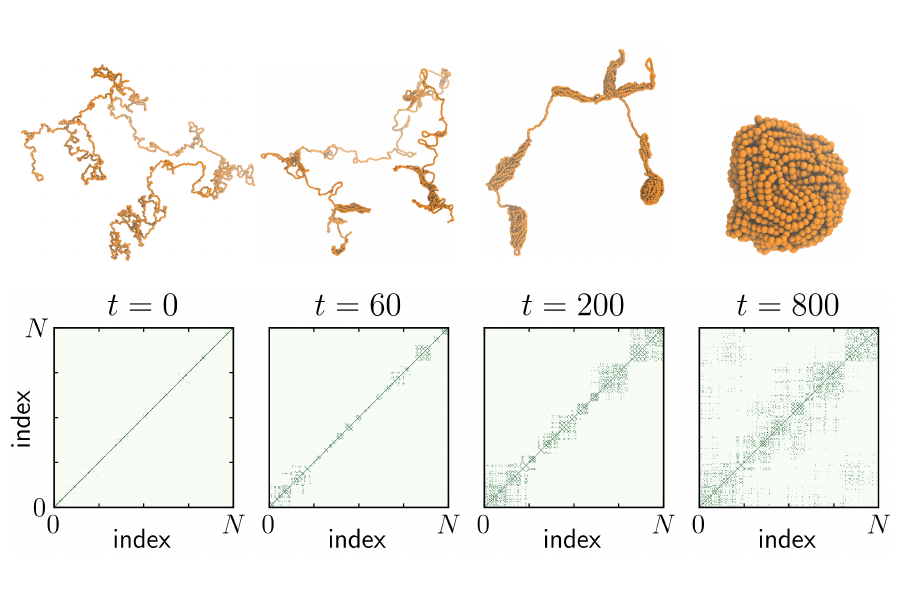}
    \caption{Collapse of a semiflexible polymer. Representative time evolution snapshots at four different times and their  corresponding contact maps during the collapse of a semi-flexible polymer with a bending stiffness $\kappa=8$ and chain length $N=2048$.}
    \label{fig:mapkappa}
\end{figure*}
\par
Next, we measure the variation in the number of clusters $N_s(t)$ of a fixed size $s$ with time, shown in Fig.~\ref{fig:fixedclusgrowthflex}. We choose the values of $s \ge 20$, which belong to the scaling regime in $\mean{C_s(t)}$ [see Fig.~\ref{fig:clusgrowthflex}(a)]. The plots include results from three different $N$ to distinctly identify the finite-size unaffected regime. The $t$ axis is scaled with $N^{1/z}$ to facilitate a better visualization of data for different $N$. At early times, mostly small clusters are present. Whereas, clusters with large $s$ appear later at the expense of the smaller ones. Thus, $N_s(t)$ for smaller $s$ starts the decay from a higher value. Also, their decay are concurrent with the rise in the number of larger clusters. 
In the intermediate to late times, data for different $s$ across different $N$ collapse on a power-law decay function. {Keeping in mind Eq.\ \eqref{eq:fvstscale}, in this regime one can quantify the data using the expression 
\begin{equation}\label{eq:fit_w}
    N_s(t)= A_w t^{-w}.
\end{equation}
The dashed line  Fig.\ \ref{fig:fixedclusgrowthflex} show the consistency of the data with this power law decay with $w=3.21$. In the next subsection, we present how the mean exponent $\mean{w}$ can be extracted from the data for both flexible and semiflexible polymers.} 

\par
Having both $z$ and $w$ in hand, we move on to investigate the distribution of cluster sizes for different times in the scaling regime and check if there exists any Vicsek-Family-like dynamic scaling. For that in Fig.\ \ref{fig:dist_cluster_flex} we present 
the scaling function $f(x)\equiv s^2 N_s(t)$ as a function of the scaling variable  $x\equiv s/t^z$ for different $t$ using data for a polymer with $N=2048$. The data for different $t$ show a reasonable data collapse. The data in the regime $x \ll 1$ show an increase consistent with the power law $\sim x^2$ as shown by the solid line, which indicates that possibly $\tau=0$. Thus one can expect that the validity of the scaling relation Eq.~\eqref{eq:sclmass}. {Recalling $z=1.67(3)$ and $w=3.21(25)$,}  it can be inferred that within numerical error for flexible polymer indeed that is the case. Overall, we conclude that colloid-like dynamic scaling for cluster-cluster aggregation can also be realized in case of clusters composed of monomers during collapse of a flexible polymer. 

\subsection{Semi-flexible polymers}\label{semiflexible}
Though pearl-necklace model describes the pathways during the collapse of a semiflexible polymer as well, the local structures of the monomer clusters are known to be different from the flexible polymers~\cite{lappala2013}. Local structures can potentially determine the shape of the clusters on polymer, which in turn can alter the kinetics of cluster aggregation. Since, we use polymers having moderate $\kappa$ values, it is ensured that the final conformation will still be a globule \cite{seaton2013flexible,majumder2021knots}, however, the dynamics of  collapse could still have lot of variations \cite{majumder2024temperature}.    
\par
 Representative evolution snapshots and corresponding contact maps during the collapse of a semiflexible polymer of length $N=2048$ with  bending stiffness $\kappa=8$ are presented in Fig.~\ref{fig:mapkappa}. At early times, there is development of clusters, leading to few contacts close to the diagonal. This is followed by aggregation of the small clusters forming of larger regions of off-diagonal contacts. Eventually, a globule is formed where long-range contacts are present over the whole chain. The progressions are similar to those in flexible polymers. However, significantly more order with diamond-like moieties can be noticed in the contact maps for the semi-flexible case. Such order has been reported earlier, and the diamond-like structures in contact maps have been attributed to the ordered local structures for semi-flexible polymers~\cite{lappala2013}.
\begin{figure}[t!]
    \centering
    \includegraphics[width=3.0in]{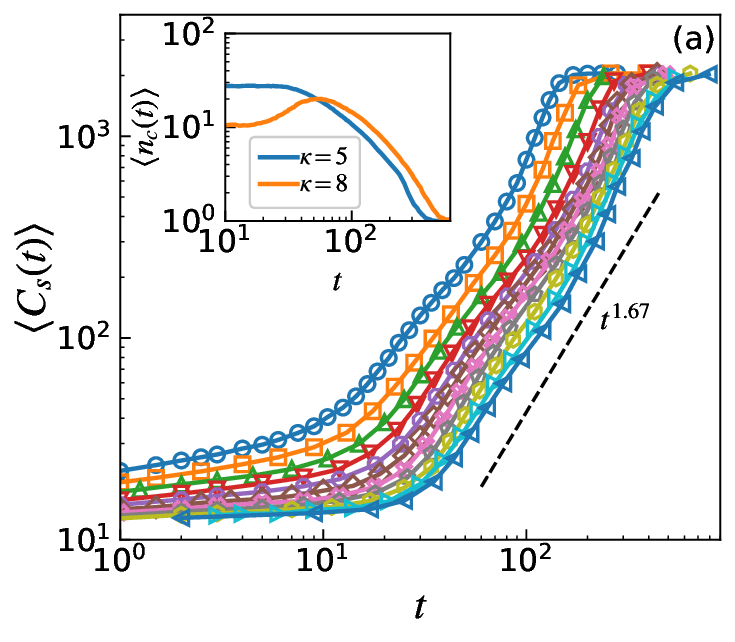}
    \includegraphics[width=3.0in]{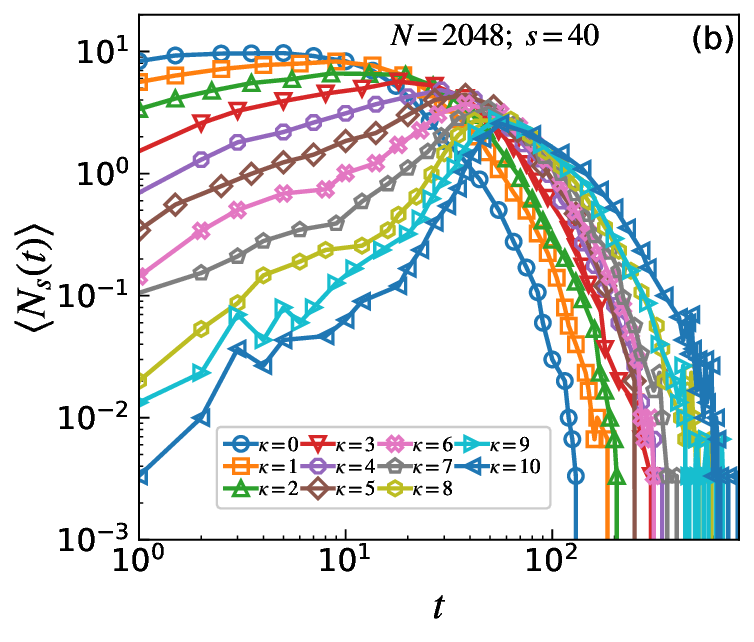}
    \caption{Cluster growth in semiflexible polymers. (a) Plot of $C_s(t)$ on a double-log scale for different $\kappa$ for a polymer with $N=2048$. The dashed line represents a power law as indicated. Note that the case $\kappa=0$ represents a flexible polymer. The inset shows the time dependence of the average number of clusters $\mean{n_c(t)}$ for two different $\kappa$. (b) Corresponding plots of the time dependence of the number of clusters $\langle N_s\rangle$ of fixed size $s=40$.}
    \label{fig:clusgrowthkappa}
\end{figure}

\par
Having an idea about the phenomenology of cluster-cluster aggregation during collapse of semiflexible polymers, we move on to quantify the cluster growth as a function of time. Corresponding plots for different $\kappa$ for a fixed chain length $N=2048$ are presented in Fig.~\ref{fig:clusgrowthkappa}(a). As in the case of flexible polymers, $\mean{C_s(t)}$ for all $\kappa$ display a relatively slow rise at early time, followed by a power-law growth in the scaling regime. As $\kappa$ increases the entry into the scaling regime gets delayed. This can be further confirmed from corresponding plots for time dependence of the number of clusters $n_c(t)$ for $\kappa=5$ and $\kappa=8$, presented in the inset of Fig.~\ref{fig:clusgrowthkappa}(a). There, the non-monotonic profile for $\kappa=8$ appears because of the time-independent size cut-off, $s_c=10$ used to identify the clusters.  If one compares the data for the cluster growth for different $\kappa$,  it appears that in the scaling regime they are almost parallel to each other, implying the presence of a universal growth exponent $z$, irrespective of $\kappa$. {Nevertheless, using the ansatz in Eq.\ \eqref{growth-ansatz} we calculate the average exponent $\mean{z}$ within the scaling regime as 
\begin{equation}\label{meanz}
\mean{z}=\left \langle \frac{d \ln \left(\mean{C_s(t)}-C_0\right)}{d \ln t}\right \rangle.
\end{equation}
Here the $\mean{\dots}$ indicates averaging over sliding time windows within the scaling regime. The obtained results are compiled in Table\ \ref{tab1} along the with their standard deviations and the used range $[t_0,t_{\rm max}]$. We have chosen $t_0$ as the point where $\mean{n_c}$ starts to decay following an initial nucleation stage. Since, the nucleation stage gets prolonged as $\kappa$ increases the chosen $t_0$s also have the same trend. This can also be appreciated from the plots of cluster growth in Fig.~\ref{fig:clusgrowthkappa}(a), which shows prolonged initial plateau with increasing $\kappa$.} The results yield no systematic dependence of the estimated $z$ values on $\kappa$. Rather it fluctuates around a mean value of $z=1.67$. Of course, to further confirm the presence of a $\kappa$-independent $z$, one can rely on finite-size scaling analysis of individual $\kappa$, as done for the flexible case in the previous subsection. However, for the sake of brevity, here, we do not present such exercises for different $\kappa$.  
\begin{table}[b!]
\caption{{Values of mean cluster growth exponent $\mean{z}$ obtained by using\ \eqref{meanz} for polymers with different stiffness $\kappa$.}}\label{tab1}
\centering
\begin{tabular}{|c|c|c|c|c|}
\hline
$\kappa$& $C_0$ & $\langle z\rangle$ & $[t_0,t_{\rm max}]$\\\hline
\hline
$0$& $32.0$ & $1.68(14)$& $[10:140]$\\
$1$& $19.0$ & $1.68(09)$& $[20:150]$\\
$2$& $18.0$ & $1.63(06)$& $[25:160]$\\
$3$& $17.0$ & $1.63(04)$& $[25:200]$\\
$4$& $16.0$ & $1.67(05)$& $[30:250]$\\
$5$& $15.0$ & $1.68(06)$& $[40:280]$\\
$6$& $8.0$  & $1.68(09)$& $[40:300]$\\
$7$& $8.0$  & $1.68(10)$& $[50:320]$\\
$8$& $6.0$ &  $1.68(13)$& $[50:330]$\\
$9$& $3.0$ &  $1.67(19)$& $[50:350]$\\
$10$&$3.0$ & $1.66(20)$& $[50:370]$\\
\hline
\hline
\end{tabular}
\end{table}
\begin{figure}[t!]
    \centering
    \includegraphics[width=3.0in]{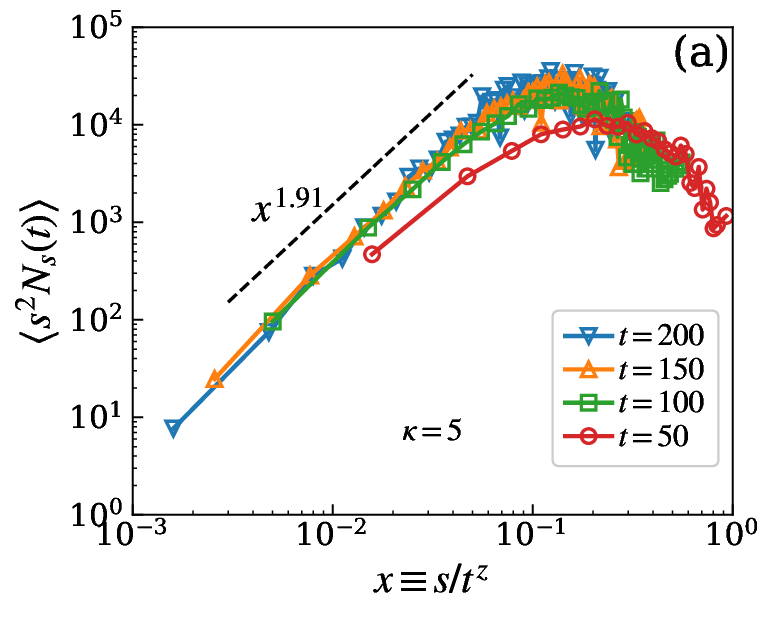}\\
    \includegraphics[width=3.0in]{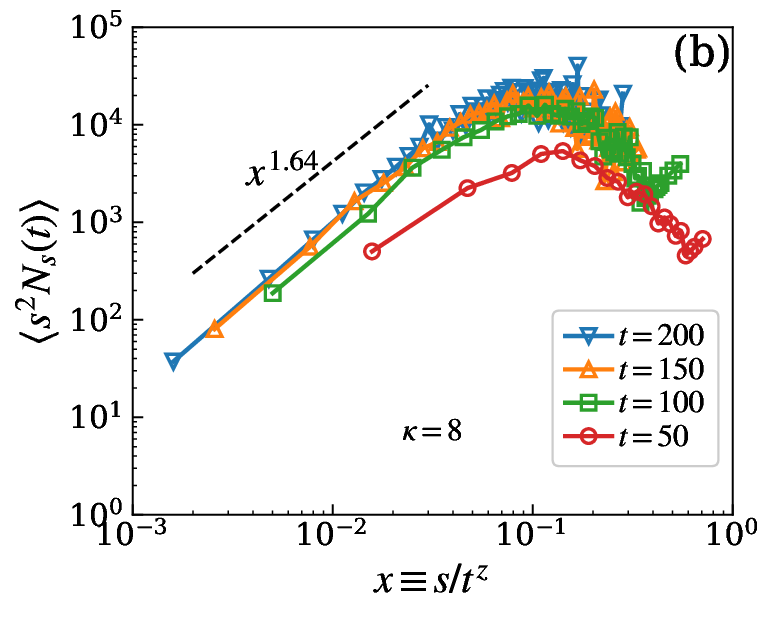}
    \caption{Verification of dynamic scaling in semiflexible polymers. Double-log plot of the scaling function $f(x)\equiv \mean{s^2N_s(t)}$ against the scaling variable $x\equiv s/t^z$ at different times for a polymer of length $N=2048$ with bending stiffness (a) $\kappa=5$ and (b) $\kappa=8$. The dashed black lines show the consistency of the data with the power law $\sim x^{\mean{\delta}}$ in the regime $x\ll1$, with respective values of $\mean{\delta}$ as indicated. Collapse of data at different times confirm the presence of colloid-like dynamic scaling in semiflexible polymers.}
    \label{fig:scaling_semi}
\end{figure}

\begin{table}[t!]
\caption{{Estimates of $\mean{w}$ using Eq.\ \eqref{meanw} to the data for decay of the number of cluster $N_s(t)$ of a fixed size $s$ as a function of time for polymers with $N=2048$ and different $\kappa$.}}\label{tab2b}
\centering
\begin{tabular}{|c|c|c|c|}
\hline
$\kappa$& $\langle w\rangle$ & $[t_0,t_{\rm max}]$\\
\hline
\hline
$0$& $3.21(25)$& $[10,140]$\\
$1$& $3.05(27)$& $[20,150]$\\
$2$& $3.26(26)$& $[25,160]$\\
$3$& $3.25(17)$& $[25,200]$\\
$4$& $3.18(13)$& $[30,250]$\\
$5$& $2.88(28)$& $[40,280]$\\
$6$& $2.78(38)$& $[40,300]$\\
$7$& $2.73(17)$& $[50,320]$\\
$8$& $2.15(11)$& $[50,330]$\\
$9$& $2.03(40)$& $[50,350]$\\
$10$&$1.56(19)$& $[50,370]$\\

\hline
\hline
\end{tabular}
\end{table}
\par
To obtain the exponent $w$, we probe the time dependence of $N_s(t)$, i.e., by calculating the variations in the count of clusters of size $s$ with time. Representative plots for $N=2048$ and $s=40$ is presented in Fig.~\ref{fig:clusgrowthkappa}(b). 
As $\kappa$ increases one notices a progressively prominent bell-shaped non-monotonic behavior for the same value of $s$. This is due to the fact that with increasing $\kappa$ nucleation of clusters to form stable clusters gets delayed, as also evident from pots of number of clusters $\mean{n_c(t)}$ in the inset of Fig.~\ref{fig:clusgrowthkappa}(a). {For each considered $\kappa$, using the expression in Eq.~\eqref{eq:fit_w} we calculate the average exponent as
\begin{equation}\label{meanw}
\mean{w}=\left \langle- \frac{d \ln N_s(t)}{d \ln t}\right \rangle.
\end{equation}
Here again the $\mean{\dots}$ represents average over sliding time windows. The obtained estimates of $w$ along with their standard deviations are quoted in Table\ \ref{tab2b} for different $\kappa$. The $t$-range used are chosen to be the same as in Table\ \ref{tab1}.} The value of $w$ appears to decrease with increasing  $\kappa$. Also, the value of $w$ becomes smaller than $2z$ as the polymer becomes stiffer. This suggests deviation from the scaling relation in of Eq.\ \eqref{eq:sclmass}. The deviations motivate us to ask whether the dynamic scaling at all exists for semi-flexible polymers. If yes, then possibly the cluster-cluster aggregation in semiflexible polymer can be represented by the other kind of dynamic scaling functional form as embedded in Eqs.~\eqref{eq:fvnomass} and \eqref{eq:sclnomass}. 
\begin{table}[t!]
\caption{{Estimates of $\mean{\delta}$ using Eq.\ \eqref{meand} to the data of the scaling function $f(x)$ as a function of the scaling variable $x$ at different times for polymers with fixed length $N=2048$ and different $\kappa$.}}\label{tab3p}
\centering
\begin{tabular}{|c|c|c|}
\hline
$\kappa$& $\langle \delta \rangle$ & $x$-range  \\
\hline
\hline
$0$&  $2.03(08)$& $[0.001,0.3]$\\
$1$&  $2.00(20)$& $[0.001,0.2]$\\
$2$&  $1.92(13)$& $[0.001,0.2]$\\
$3$&  $2.06(06)$& $[0.001,0.15]$\\
$4$&  $1.94(24)$& $[0.001,0.12]$\\
$5$&  $1.91(13)$& $[0.001,0.08]$\\
$6$&  $1.77(19)$& $[0.001,0.07]$\\
$7$&  $1.74(12)$ & $[0.001,0.06]$\\
$8$&  $1.64(10)$& $[0.001,0.05]$\\
$9$&  $1.57(25)$& $[0.001,0.04]$\\
$10$& $1.48(28)$& $[0.001,0.03]$\\

\hline
\hline
\end{tabular}
\end{table}

\begin{figure}[b!]
    \centering
    \includegraphics[width=3.0in]{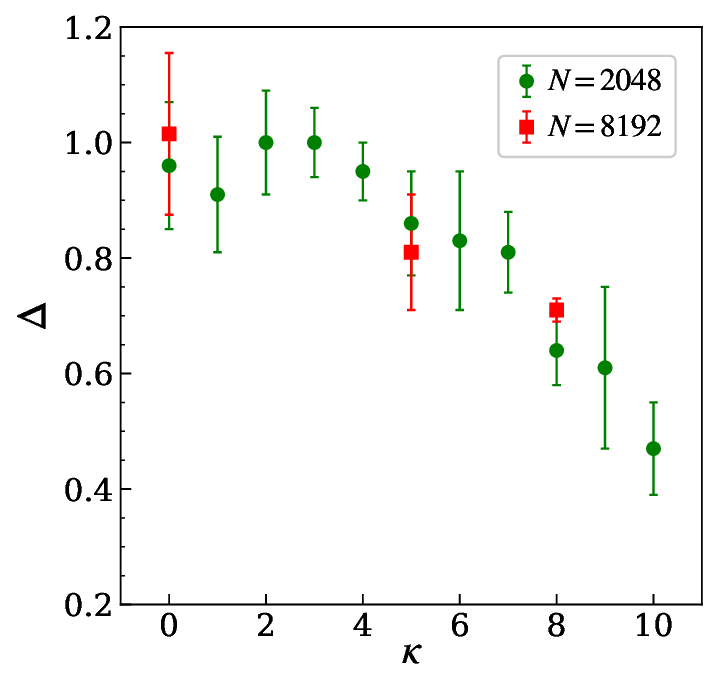}
    \caption{Deviation from the Vicsek-Family-like scaling. Plot of $\Delta$ accounting the deviation from scaling relations in Eq.\ \eqref{eq:sclmass}  as a function of the bending stiffness. For $N=8192$, results from only three different values of $\kappa$ are shown.}
    \label{fig:VF_scaling_deviation}
    \end{figure}
\par
Particularly for the scaling relation in Eq.~\eqref{eq:sclnomass}, the exponent $\tau$ corresponding to the cluster size distribution becomes important. To extract the exponent, as in flexible polymer (see Fig.\ \ref{fig:dist_cluster_flex}), we first verify if the dynamic scaling holds in semiflexible polymers as well by using the data for cluster-size distributions at different times during the collapse. Our exercise reveals that indeed the dynamic scaling holds for semiflexible polymers. In Figs.\ \ref{fig:scaling_semi}(a) and (b) we illustrate this finding for two $\kappa$ values for a polymer of length $N=2048$. There the plots of the scaling function $f(x)\equiv \mean{s^2 N_s(t)}$ at different times as a function of the scaling variable  $x\equiv s/t^z$ show reasonably good data collapse for both $\kappa$. For other values of $\kappa$ as well we obtain such data collapse. Note that the times used in (a) and (b) are different because as $\kappa$ increases the entry to the scaling regime gets delayed, as already discussed. In fact, for $\kappa=8$ one can notice that the data for $t=50$ and $100$ are a bit off for the same reason. In the region $s/t^z \ll 1$, the data show a power-law behavior with an expected exponent $\delta = 2-\tau$, as discussed earlier. {We observe a consistent behavior in this regime, and therefore consider the power law
\begin{equation}\label{fit_delta}
 f(x)=A_{\delta} x^{\delta},
\end{equation}
in this regime to calculate the average exponent as 
\begin{equation}\label{meand}
\mean{\delta}=\left \langle- \frac{d \ln f(x)}{d \ln x}\right \rangle,
\end{equation}
where the $\mean{\dots}$ now corresponds to a sliding $x$ window within the regime $x \ll 1$.
The estimates obtained and the used $x$-range are quoted in Table\ \ref{tab3p}.} The results indicate that $\delta$ decreases monotonically as a function of $\kappa$, i.e., $\tau$ increases monotonically with $\kappa$. Now, recalling the estimated values of $z$ and $w$, reveals that as $\kappa$ increases one deviates from the relation in Eq.\ \eqref{eq:sclmass}. The deviation can be quantified by $\Delta$ as a function of $\kappa$, where 
\begin{equation}\label{eq_S1}
\Delta=\frac{w}{2z} 
\end{equation}
is a measure of the deviation from scaling relations in Eq.~\eqref{eq:sclmass}. The corresponding plots as a function of $\kappa$ is presented in Fig.\ \ref{fig:VF_scaling_deviation}. The data clearly depict that for small $\kappa$ the scaling relation in Eq.\ \eqref{eq:sclmass} holds true with $\Delta \approx 1$ and for $\kappa \ge 5$ it does not hold anymore. We have also checked the deviation from Eq.\ \eqref{eq:sclnomass}, however, there we do not find any meaningful trend.  


\par
In cluster-cluster aggregation of purely particle systems, the validity of the scaling functions in Eqs.\ \eqref{eq:fvnomass} or \eqref{eq:fvmass} depends on how the clusters diffuse before they coalescence with each other \cite{meakin1985}. The diffusion constant of {cluster depends} on the size $s$ as described in Eq.\ \eqref{D_eq}. As already pointed out, there exists a critical value of the exponent $\gamma_c$, such that for $\gamma > \gamma_c$ one realizes the scaling relation in Eq.\ \eqref{eq:sclnomass} and for $\gamma < \gamma_c$ the relation in Eq.\ \eqref{eq:sclmass} holds true. In the present case, as in most of realistic situations, obviously $\gamma < 0$, which  indicates that $D_s$ should decrease with increasing $s$. In addition, our results for small $\kappa$ value also suggests that they belong to the scenario of $\gamma < \gamma_c$. To dig further about the deviation of the data from the scaling relation for $\kappa \ge 5$, one can rely on a comparative analyses of the diffusion constant $D_s$ as a function of $\kappa$. 

\par
For estimating the $D_s$ one needs to monitor a cluster of reasonably large size, and from its trajectories needs to calculate mean-squared displacement of the center of mass of the cluster. However, during the collapse of a polymer the clusters formed are not sufficiently long lived to have a trajectory that can provide a reasonable estimate of $D_s$. Hence, we opt for an indirect comparison of $D_s$ for different $\kappa$. We recall that the local structures in semiflexible polymers are more order than those in a flexible polymer.  Differences in order may lead to changes in shape. Thus, a cluster of size $s$ can have different hydrodynamic or Stokes-Einstein radius $R_h$, depending on the stiffness of the polymer, resulting in variation in effective diffusion coefficient as $D_s \sim 1/R_h$.
\begin{figure}[t!]
    \centering
    \includegraphics[width=3.0in]{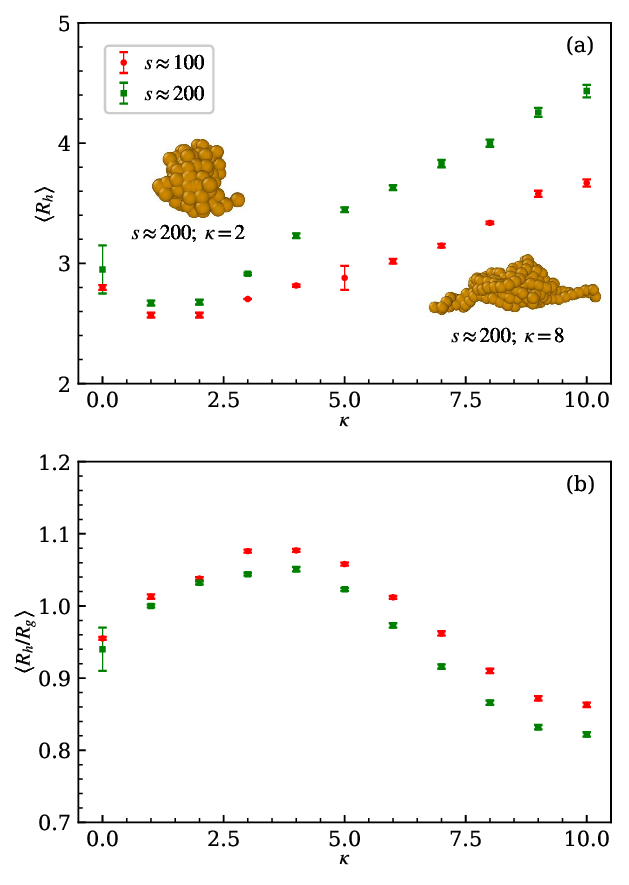}
    \caption{Hydrodynamic radius of clusters. (a) Average hydrodynamic radius $\langle R_h\rangle$ of clusters as a function of $\kappa$ for two different choices of cluster sizes $\langle s\rangle$ for a polymer of length $N=2048$. The snapshots represent typical clusters of size $\langle s\rangle \approx 200$ obtained at a small and a large $\kappa$, as indicated. (b) Corresponding variation of the ratio $\langle R_h/R_g \rangle$, where $R_g$ is the radius of gyration of the cluster.}
    \label{fig:hydro_radius_dimension}
\end{figure}

\par
To test the hypothesis, we estimate $R_h$ of clusters of fixed size $s \approx 100$ and $200$, which represent cluster sizes well within the scaling regime for all $\kappa$. For the calculation, we use the simplified Kirkwood approximation given as \cite{kirkwood1954,liu2003,pesce2023,chakraborty2025}
\begin{equation}\label{Rh_eqn}
    \frac{1}{R_h} = \frac{1}{s}\sum_{i=1}^{s}\sum_{j=1,j\ne i}^{s}\left \langle \frac{1}{d_{ij}}\right \rangle,
\end{equation}
where $d_{ij}=|\vec{r}_i-\vec{r}_j|$ is the distance between any two monomers $i$ and $j$ within the cluster of size $s$. Note that here $\langle \dots \rangle$ represents averaging over different clusters. Along the same line we also calculate the radius of gyration of the cluster as 
\begin{equation}\label{Rg_eqn}
R_g=\sqrt{\frac{1}{2s^2}\sum_{i,j}(\vec{r}_i-\vec{r}_j)^2}.
\end{equation}
The ratio of $R_h$ and $R_g$ gives a comparative idea about the shapes of the clusters of same size. 
The variation of $R_h$ with $\kappa$ is shown in  in Fig.~\ref{fig:hydro_radius_dimension}(a) for the two values of $s$. The profiles show a marginal dip around $\kappa=2$ following which it increases steadily. This suggests that clusters of same size will be more elongated and loosely bound as $\kappa$ increases, as also revealed from the typical clusters of size $s\approx 200$ for a small and large $\kappa$. {Similarly, in Fig.\ \ref{fig:hydro_radius_dimension}(b)} the ratio $R_h/R_g$ indicate a change in shape of the clusters around $\kappa \approx 5$. Smaller value of the ratio indicates a smaller fractal dimensions of the clusters with increasing $\kappa$. This the behavior of the data in Fig.\ \ref{fig:hydro_radius_dimension} strongly implies a change in the diffusion constant $D_s$ with $\kappa$. A comparison with Fig.\ \ref{fig:VF_scaling_deviation} confirms that the deviation from the scaling relation in Eq.~\eqref{eq:sclmass} onsets around the same value of $\kappa$. Since the scaling relation in Eq.\ \eqref{eq:sclmass} is based on the diffusion of the clusters and its dependence on its size, this correlation is certainly strong. Here, we confirm that indeed there is a change in the diffusion of the clusters at around the same value of $\kappa$ where Eq.\ \eqref{eq:sclmass} breaks down. Of course, a detailed theoretical treatment of this correlation is needed to further substantiate it.  
\begin{figure}[t!]
    \centering
    \includegraphics[width=3.0in]{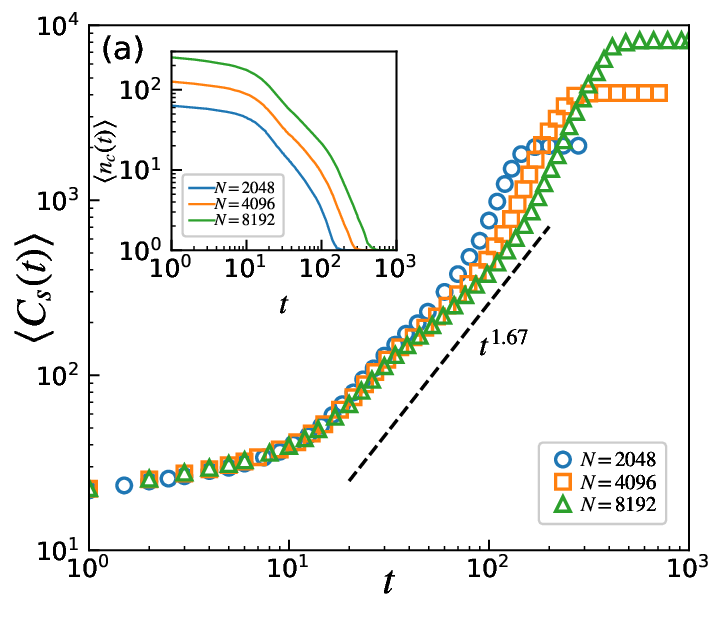}\\
    \includegraphics[width=3.0in]{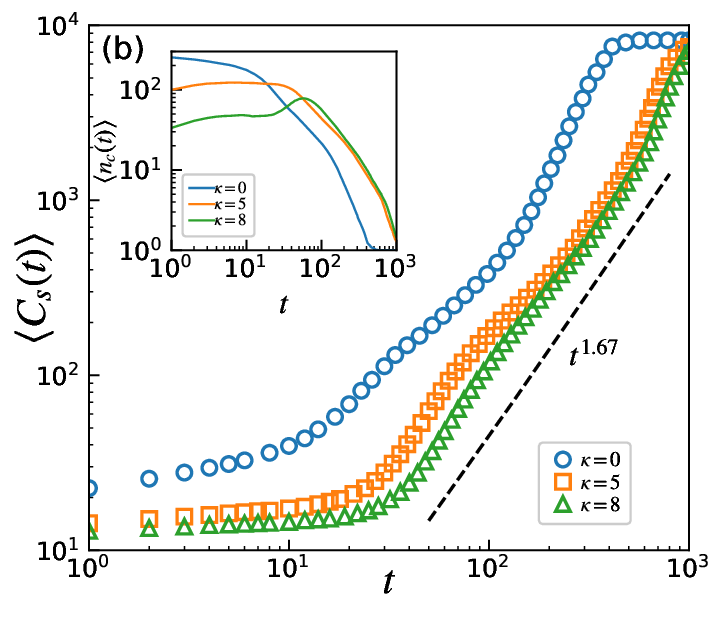}
    \caption{Cluster growth for longer chins. (a) Cluster growth for flexible polymer with three different $N$. The inset shows the corresponding time dependence of the number of clusters $\mean{n_c(t)}$.  (b) Corresponding variation of the cluster growth for three different $\kappa$ for the longest chain with $N=8192$. The inset presents the corresponding plots for $\mean{n_c(t)}$. The dashed lines in the main plots represent the same power law $\sim t^{1.67}$.}
    \label{clgrowth_long}
\end{figure}
{
\subsection{Results for longer chains}\label{long_chains}
Here we present results for a longer chain with $N=8192$ for a flexible polymer ($\kappa=0$) and two semiflexible polymers with $\kappa=5$ and $8$. The results presented here re-establish the claims made using relatively shorter chains. We start with presenting the cluster growth in Fig.\ \ref{clgrowth_long}. For a comparison, in Fig.\ \ref{clgrowth_long}(a) we show the data for the flexible polymer along with relatively shorter chains $N=4096$ and $2048$. Data for all system sizes follow each other until the shorter ones start deviating from the larger ones at late time before they reach the finite-size limit. This deviation is a virtue of the effect that at late times when the clusters are considerably huge, the number of clusters are limited and the tension in parts of the chain connecting them slowly becomes maximum. This eventually leads to a faster coalescence of the cluster than purely diffusive motion. This is in accordance with phenomenology described by Klushin \cite{klushin1998}. Thus, for long enough chains the overall collapse possibly starts with the pearl-necklace phenomenology before crossing over to a chain-tension mediated coalescence, described by Klushin. A view of the full trajectory for longer chain in confirms this speculation. For small chain length, the duration of the latter stage is negligible before attaining the finite-size limit. The time dependence of the number of clusters, shown in the inset, also bears this signature with a steep fall at late times for longer chains. 
\begin{figure*}[t!]
    \centering
    \includegraphics[width=6.0in]{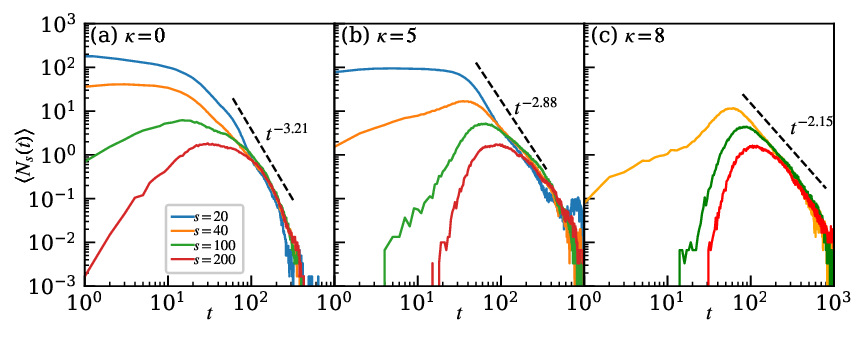}\\
        \caption{Time dependence of the number of clusters of fixed sizes $N_s(t)$ for a long polymer with (a) $\kappa=0$, (b) $\kappa=5$ and (c) $\kappa=8$. The dashed lines show consistency of the data to the power-law decays obtained using a shorter chain.}
    \label{fixed_clgrowth_long}
\end{figure*}
\par
The same observation can also be made from the cluster growths of polymers with $\kappa=5$ and $8$ for $N=8192$, as presented in Fig.\ \ref{clgrowth_long}(b). A careful observation of the trajectories for all three $\kappa$ will reveal that before the collapse reaches the late chain-tension mediated stage, the clusters are moving more or less in a Brownian path, resulting in a diffusive transport mechanism as observed in colloidal systems. However, the emergence of the late chain-tension mediated phase makes this diffusive scaling regime limited. For larger $\kappa$ one may expect that these regime may be further limited by the stiffness of the backbone and thereby effectively larger Kuhn lengths \cite{de1979scaling}. However, since the considered $\kappa$ are in the moderate range  ensuring that the final state is still a collapsed globule, this effect is perhaps not that magnified. On the other hand, within this moderate range the effective diffusive constant of the clusters decreases due to the increased $\kappa$, which affects only the amplitude of the cluster growth, and not the exponent as in other particle coarsening system \cite{majumder2013temperature}. This smaller growth amplitude, consequently, prolongs the diffusive cluster-coalescence regime as evident from the plots in Fig.\ \ref{clgrowth_long}(b). Thus, results from polymer chains of length $N=2048$ were sufficient to focus on the cluster coalescence in the diffusive regime. In any case, for $N=8192$ also we have extracted the corresponding exponents within the scaling regime of interest.

\par
For the cluster growth of the chain with $N=8192$, the obtained exponents are $\mean{z}=1.69(13)$, $1.67(19)$ and $1.64(02)$, respectively, for $\kappa=0$, $5$ and $8$, which is consistent with $z\approx 1.67$ quoted for the shorter chain with $N=2048$. Similarly, for the time-dependent decay of number of clusters of fixed size $N_s(t)$ are also consistent with the behavior claimed for a shorter chain, as shown in Fig.\ \ref{fixed_clgrowth_long}. The corresponding estimated values of $\mean{w}$ are $3.43(42)$, $2.69(06)$ and $2.33(02)$. Finally, in all the cases we have realized the dynamic scaling proposed in \eqref{eq:fvnomass} and \eqref{eq:fvmass}, as illustrated in Fig.\ \ref{sc_fx_long}. The corresponding estimated $\mean{\delta}$ are, respectively, $1.96(15)$, $1.60(17)$ and $1.57(07)$. Using these values of the exponents, we have also determined the deviation parameter $\Delta$ using Eq.\ \eqref{eq_S1}. This calculated $\Delta$ for $N=8192$ for three values of $\kappa$ are also shown in Fig.\ \ref{fig:VF_scaling_deviation}, which within the error bars are in accordance with the data for $N=2048$.
\begin{figure*}[t!]
    \centering
    \includegraphics[width=6.0in]{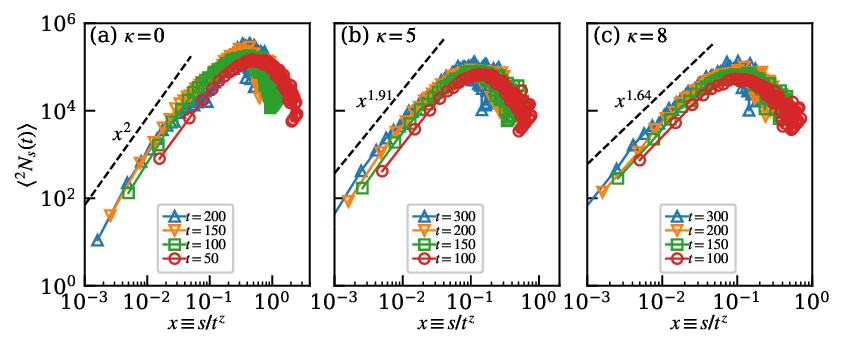}\\
        \caption{Illustration of the dynamic scaling for a longer chain with (a) $\kappa=0$, (b) $\kappa=5$ and (c) $\kappa=8$. The dashed lines illustrate the consistency of the data with the respective power-laws obtained using a shorter chain in the regime $x\ll1$.}
    \label{sc_fx_long}
\end{figure*}
}
\section{Discussion}\label{conclusion}
We have investigated the collapse kinetics for a range of homopolymers from flexible to moderately stiff ones, using MD simulations at constant temperature of a bead-spring model. Following a quench from an extended coil-like conformation to a temperature below the collapse transition we have monitored the trajectory of the polymer until it becomes a globule. The nonequilibrium pathway of the collapse more or less follows the popular pearl-necklace picture, irrespective of the bending stiffness $\kappa$. The intermediate to late stage of the pathway is dominated by the aggregation of ``pearls'' or clusters of monomers. Drawing analogies with the cluster formation during colloidal assembly,  here, we have aimed to verify the existence of associated dynamic scaling of cluster-cluster aggregation during polymer collapse. 
\par
The verification was performed by measuring the time dependence of the average cluster size $C_s(t)$, time dependence of the number of cluster of fixed size $N_s(t)$ and the cluster-size distribution at different times. Our results for the cluster growth indicates a universal power-law behavior $C_s(t) \sim t^z$ with $z\approx 1.67$, irrespective of $\kappa$. However, the exponent $w$ associated with the long-time decay $N_s(t) \sim t^{-w}$, varies inversely with $\kappa$. Interestingly, distributions of cluster sizes exhibit the dynamic scaling of the form $N_s(t)\sim s^{-2} f(s/t^z)$, irrespective of $\kappa$.  Otherwise difficult to estimate from our data, the realization of this scaling also allowed us to calculate the the exponent $\tau$ associated with the cluster-size distribution at a fixed time as $N_s \sim s^{-\tau}$. From the estimated values of $z$, $w$ and $\tau$ we have discovered that for small $\kappa < 5$, in addition to the dynamic scaling, the relation $w=2z$ is also obeyed.  For $\kappa \ge 5$ the estimated exponents show systematic deviation from the relation as a function of $\kappa$. This suggests that even though presence of dynamic scaling can be realized during polymer collapse, irrespective of the bending stiffness, the nature of scaling varies in terms of the relation between the dynamic exponents.
\par

The dynamic scaling, observed here for polymers is routinely observed for a wide range of colloidal aggregations. There, observation of a specific kind of scaling function is determined by factors such as size dependent diffusion constant of the clusters. In polymers, larger $\kappa$ lead to higher local order and different shape of the clusters compared to flexible or weakly stiff polymers. Thus, to find the origin of the breakdown of the scaling relation for large $\kappa$ in polymer collapse, we have estimated the hydrodynamic $R_h$ of  the \enquote{pearls} or clusters. For the same monomer count $s$, the clusters belonging to semiflexible polymers show larger $R_h$, implying a smaller effective diffusion coefficient. The $\kappa$-dependent profile of $R_h$ follows the same trend as the scaling relation deviates from flexible polymers to semiflexible polymers. Thus, we speculate that the local order induced shape change as the cause of the change in the governing scaling law in cluster-cluster aggregation during polymer collapse. {To further confirm our findings we have also presented results for a polymer with a much longer chain length $N=8192$, albeit, for only three choices of $\kappa$. There, in addition to the pearl-necklace collapse, we found signature of a chain-tension mediated cluster growth at very late stage when the cluster are much larger. However, before that stage the cluster growth follows the same dynamic scaling as presented for the shorter chains. In future it would be interesting to investigate the scaling behavior of the final chain-tension mediated stage in detail, for which one needs to simulate even longer chains requiring more computational efforts. We propose this as a future task to explore.}
\par
Historically, tracking the collapse of single polymers experimentally was difficult due to technical challenges in generating ultra-dilute solutions or obtaining sufficiently long mono-dispersed polymers~\cite{nishio1979first, chu1995}. Significant progress in addressing these issues have been achieved through high resolution of modern instruments, e.g., small-angle X-ray scattering, single-molecule fluorescence, and dielectric spectroscopy have enabled probing of conformational changes in individual macromolecule with time~\cite{pollack2001,Sadqi2003,tress2013,hofmann2012,naudi2021}. However, validation of the dynamic scaling behavior observed in this work would require well-resolved visualization of the intermediate states with clusters along the chain. While time-resolved investigations of colloidal aggregation and peptide assembly are now routinely performed~\cite{grober2023, lei2017, newcomb2012,chakraborty2019}, capturing the rapid collapse dynamics of polymers remains challenging due to spatio-temporal limitations of the conventional imaging techniques such as bright-field, confocal, time-resolved atomic force, or cryogenic transmission electron microscopy (cryo-TEM). Notably, the collapse time is expected to scale with polymer length as $\tau_c \sim N^{\nu}$, where typically $\nu=1-2$ \cite{majumder2020}.  Thus, recent advances in synthesis of ultra-high-molecular-weight polymers have the potential to extend collapse times to the scale of hours~\cite{gao2021, chou2024, eades2025}. Additionally, viscosity and temperature of the ambient solvent can be tuned to prolong the collapse time~\cite{majumder2017, majumder2024temperature}. Under such conditions and in sufficiently dilute solutions, cryo-TEM may become a viable technique for imaging and quantifying the size distribution of intermediate clusters at various time points~\cite{newcomb2012,patterson2017}. Although such experimental efforts would require substantial resources and optimization, we anticipate that continued progress in polymer synthesis and imaging technologies will soon enable the direct observation of these transient collapse intermediates. 
\par
For colloids, these scaling principles are known to be universal, independent of the composition of colloids~\cite{lin1989universality,lin1990universal}. The observation of similar principles in polymer expands the domain of the universality of these theories. The analogies with colloidal aggregations further reaffirms the validity of droplet coalescence model of polymer collapse. As a next step, it would be worth to verify the same for collapse of heteropolymers or protein-like heteropolymers. {In this regard, for heteropolymers with a Gaussian and uniform distributions of polydispersity, the collapse dynamics is found to be universal, irrespective of the degree of polydispersity \cite{singh2022universality}. Albeit, it shows significant differences in the final monomer re-organization stage, which is crucial in achieving the final native state.}
A step even further would be to perform all-atom MD simulations of polypeptide collapse and check the universality of the dynamic scaling. In this regard, one can also investigate the collapse of a real globular protein.


\acknowledgments
The work was funded by the Anusandhan National Research Foundation (ANRF), Govt.\ of India through a Ramanujan Fellowship (File no.\ RJF/2021/000044). The authors thank Aritra Sarkar for useful discussions on the experimental methods.
%
\end{document}